# Electrically Switchable Metasurface for Beam Steering Using PEDOT

*Juliane Ratzsch[1], Julian Karst[1], Jinglin Fu[1], Monika Ubl[1], Tobias Pohl[1], Florian Sterl[1], Claudia Malacrida[2], Matthias Wieland[2], Bernhard Reineke[3], Thomas Zentgraf[3], Sabine Ludwigs[2], Mario Hentschel[1], and Harald Giessen[1]*

[1] 4th Physics Institute and Research Center SCoPE, University of Stuttgart, Pfaffenwaldring 57, 70569 Stuttgart, Germany

[2] IPOC-Functional Polymers, Institute of Polymer Chemistry, University of Stuttgart, Pfaffenwaldring 55, 70569 Stuttgart, Germany

[3] Department of Physics, Paderborn University, 33098 Paderborn, Germany

Corresponding Author:

Prof. Dr. Harald Giessen

4[th] Physics Institute, University of Stuttgart

E-Mail: h.giessen@pi4.uni-stuttgart.de

Tel: +49 711 685 65111

Abstract

Switchable and active metasurfaces allow for the realization of beam steering, zoomable metalenses, or dynamic holography. To achieve this goal, one has to combine high-performance metasurfaces with switchable materials that exhibit high refractive index contrast and high switching speeds. In this work, we present an electrochemically switchable metasurface for beam steering where we use the conducting polymer poly(3,4-ethylene-dioxythiophene) (PEDOT) as an active material. We show beam diffraction with angles up to 10° and change of the intensities of the diffracted and primary beams employing an externally applied cyclic voltage between -1 V and +0.5 V. With this unique combination, we realize switching speeds in the range of 1 Hz while the extension to typical display frequencies in the tens of Hz region is possible. Our findings have immediate implications on the design and fabrication of future electronically switchable and display nanotechnologies, such as dynamic holograms.

Introduction

The manipulation of optical wavefronts via lenses, mirrors, or any other optical devices is present in all our everyday lives. This includes, e.g., bulky lenses made from glasses, Fresnel lenses in smartphone illumination optics, or spatial light modulators and liquid crystals in displays. However, emerging display technologies such as dynamic holography or augmented and virtual reality require ever-increasing pixel densities and thus new smart optical methods with the ability to manipulate optical wavefronts, beam paths, polarization, or similar actively on ultra-small length-scales [1–3]. For the realization, the research field of plasmonics has gained significant interest over the last years as plasmonic nanoparticles allow to focus, manipulate, or steer light on the nanometer scale at subwavelength dimensions [4–6]. The combination of optically active and externally switchable materials with plasmonics into hybrid nanosystems has increased their applicability even further and opened the door towards active plasmonically-driven light manipulation [7]. Especially, plasmonic metasurfaces, that are, artificial sheet materials with sub-wavelength thickness, allow the realization of flat optical components with unique optical properties [8–13]. Furthermore, the combination with phase change materials inaugurates active plasmonic optical applications such as active beam switching [14,15], zoom lensing[16], dynamic holography [17], dynamic plasmonic color displays [18,19] and many other. One possibility to realize these hybrid nanosystems is the fabrication of nanoparticles directly from phase change materials such as magnesium [20–23], palladium [24–27], and yttrium [28]. On the other hand, phase change materials such as polyaniline (PANI) [29,30], germanium–antimony–tellurium-based film films (GST) [31–34], liquid crystals [35,36], or $VO_2$ [37,38] are widely used to combine with nanoparticles to allow active switching. However, widespread and commercial applications are, so far, hindered by several limiting factors such as material degradation, slow switching speeds, low optical contrast (refractive index shifts), and similar. Even more, most material phase transitions are only accessible via variations in, e.g., temperature or exposed gases and not electrically.

Here, we present an optically active system to perform switchable beam steering realized via a novel hybrid metasurface. It consists of a unique combination of gold plasmonic nanoantennas and the electrically switchable and conducting polymer poly(3,4-ethylenedioxythiophene) (PEDOT) [39]. We show that tuning of geometric parameters allows switching angles and beam diffraction of up to 10°, while the efficiency of the hybrid metasurface and thus the intensity of diffracted light can be actively controlled via an externally applied cyclic voltage between only -1 V and +0.5 V. An increase of the scan-rate of the cyclic voltammetry reveals switching frequencies close to 1 Hz while an extension to typical display frequencies in the tens of Hz range is possible, only limited by the measurement components.

Experimental Details

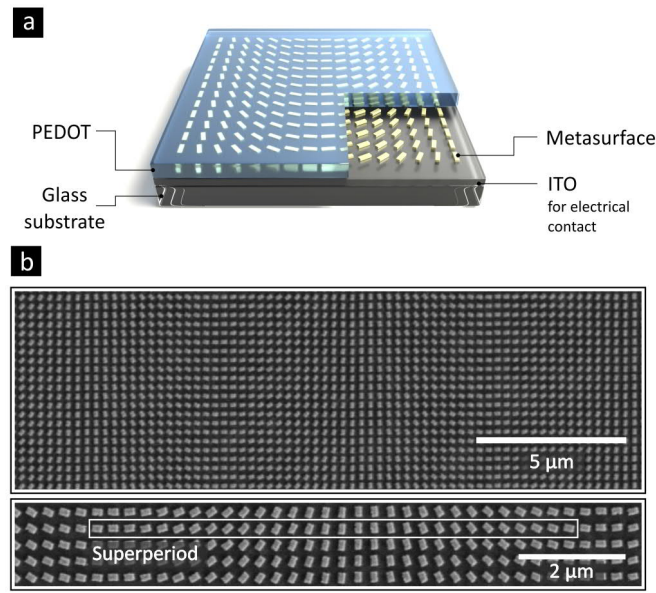

Figure 1. (a) Schematic picture of the active plasmonic metasurface for beam steering. Gold nanoantennas on an ITO covered glass substrate, coated with electropolymerized PEDOT, which acts as the reversibly switchable layer (refractive index shift) during cyclic voltammetry. ITO is used for electrical contact. (b) SEM image of the metasurface. The superperiod, consisting of 30 antennas, is marked in white in the lower zoom-in image.

In order to design the plasmonic nanostructures, we used the transient solver of CST Microwave Studio Suite in a wavelength range between 600 nm and 1200 nm. Figure 1a shows the material stack and geometries of the simulated system. We simulated a single unit-cell with open boundary conditions in the z-direction to emulate free space and periodic boundary conditions in x- and y-direction to simulate an infinite array of antennas. To optimize the on/off behavior, we performed two simulations—one series with the refractive index of the reduced PEDOT and one sequence with the oxidized PEDOT. With a parameter scan, the geometry was optimized to find an antenna geometry, which leads to a significant shift of the plasmonic resonance when the refractive index of the PEDOT changes. Resulting amplitude and phase spectra of the transmission coefficient for cross- and co-polarization can be found in Figure S1 in the Supplementary Information. The resonance shift allows us to turn on and off the anomalous refraction of the plasmonic antenna array at a specific wavelength. The refractive index of PEDOT was taken from Stockhausen et. al. [39]. Further, the refractive index of gold was taken from Yakubovsky et. al. [40].

The final metasurface consists of gold nanoantenna arrays with progressively rotated elements along one axis [41], as shown in the SEM images in Figure 1b. Fabrication is done via Electron-beam lithography (Raith eLine Plus). The antennas have a size of 200 nm x 135 nm, a thickness of 50 nm, and a period of 300 nm in x- and y-direction and are placed on a glass substrate covered with a thin layer (20 nm) of Indium-Tin-Oxide (ITO) for electrical contact. Adjacent antennas are rotated with an incremental rotation angle of 6°. The resulting superperiod after 180° rotation is 9 µm and comprises 30 antennas, as marked in the lower SEM image in Figure 1b. When illuminating this metasurface with circularly polarized light of one handedness, the design results in an additional transmitted beam that is diffracted from the main beam by several degrees and possesses opposite circular polarization. To make our hybrid metasurface electrically switchable in the visible spectral range, we cover the metasurface with a 150 – 200 nm thick poly(3,4-ethylenedioxythiophene) (PEDOT) layer which is deposited by electropolymerization [42,43]. This PEDOT layer undergoes a significant refractive index change in the visible spectral range when the applied voltage is varied by cyclic voltammetry [44] and thus results, in combination with the metasurface, in electrochemically activated switchable beam steering. Please note that modifying the deflection angles or similar would require an active change of the metasurface parameters (antenna geometry, spacing, etc.) or the combination of multiple metasurfaces. So far, this is not possible with our hybrid metasurface design. See Figure S2 in the Supplementary Information for details on the electrodeposition and structural change of PEDOT.

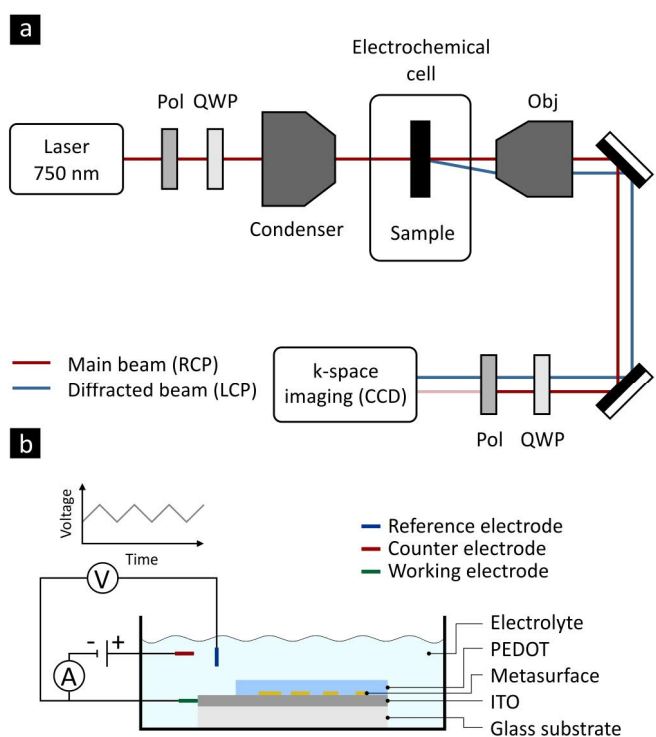

Figure 2. (a) Schematic picture of the k-space setup to measure the diffraction of the actively switchable metasurface. The main incident beam is right-circularly polarized (RCP), while the diffracted beam left-circularly polarized (LCP). The circular analyzer (second pair of QWP and Pol) allows for diminishing the main beam. (b) Schematic picture of the electrochemical cell where the PEDOT is actively switched via cyclic voltammetry.

A schematic drawing of our main setup to perform angle-resolved imaging is shown in Figure 2a. It consists of a modified transmission microscope (Nikon Eclipse TE2000-U) in combination with a home-built k-space imaging module [27]. We use a tunable laser (NKT Photonics SuperK Extreme) as an illumination source, set at a wavelength of λ = 750 nm. To obtain right-circularly polarized (RCP) light we use a polarizer (Thorlabs LPVIS100) and a broadband quarter-wave plate (QWP) (B. Halle RAC 5.4.20). As we image in k-space, the condenser is set highly defocused to allow for the best possible normal incidence on the metasurface. The metasurface is placed at the sample position in the front focal plane of the objective (Nikon Plan Fluor ELWD 40x, N.A. 0.60), which collects the main transmitted and the diffracted beam. The diffracted beam is left-circularly polarized (LCP) and thus possesses opposite circular polarization compared to the main beam. Consequently, a circular analyzer consisting of another QWP (B. Halle RAC 5.4.20) and polarizer (Thorlabs LPVIS100) allows for the adjustment of the intensity of two beams. As in our experiments the main beam is more intense, we use this circular analyzer to attenuate the main beam resulting in similar intensities of both beams before starting the switching experiment. To be able to resolve the diffraction angles of the two beams, we directly image the objective back focal plane (k-space) with a modified 4-f setup [45] on a CCD camera (Allied Vision GC2450C). As schematically depicted in Figure 2a, the sample is placed inside a custom-made electrochemical cell to perform cyclic voltammetry on the active PEDOT layer with a three-electrode setup. A schematic drawing of the electrochemical setup is shown in Figure 2b. As electrolyte we use a 0.1 mol/L $CH_3CN/Bu_4NPF_6$ solution. We attach the working electrode (platinum (Pt) wire) to the electrically conducting and partially uncovered ITO layer underneath the metasurface. The counter electrode (Pt wire) and reference electrode (silver chloride silver wire, Ag/AgCl) are in contact with the electrolyte solution. The reference electrode is placed near the working electrode to minimize Ohmic drops of the voltammogram. Finally, the three electrodes are connected to a potentiostat (BioLogic SP-200) to perform cyclic voltammetry between -1 V and +0.5 V vs. Ag/AgCl. The typical voltammogram of the PEDOT deposited on our metasurface as well as a ferrocene reference is shown in Figure S2 in the Supplementary Information.

## Results and Discussion

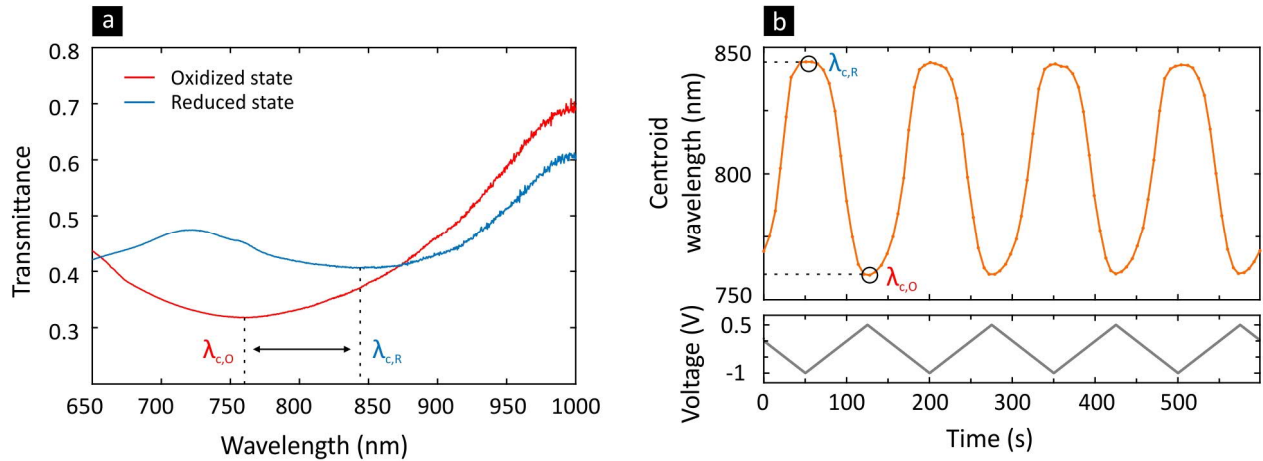

Figure 3. Spectral shift of the plasmonic resonance of the hybrid metasurface. (a) Transmittance of the metasurface in the oxidized (red) and reduced (blue) state of PEDOT at the extrema during cyclic voltammetry (oxidized and reduced PEDOT) (b) Centroid wavelength and applied voltage over time. The voltage is measured vs. a Ag/AgCl pseudoreference electrode. Marked are $\lambda_{c,O}$ of the oxidized and $\lambda_{c,R}$ of the reduced state of PEDOT.

The effect of beam steering is obtained via the refractive index shift of the PEDOT layer during cyclic voltammetry. This change in refractive index causes the plasmonic resonance of the metasurface to shift and thus the efficiency of the beam diffraction to vary. The spectral response of our hybrid metasurface is shown in Figure 3a. For spectral measurements, we replace the k-space imaging module in Figure 2a with a grating spectrometer (Princeton Instruments SP2500i) and a Peltier-cooled front-illuminated CCD camera (Acton PIXIS 256E). In the oxidized state (red curve), we find a broad plasmonic resonance with a centroid wavelength around $\lambda_{c,O}$ = 760 nm. In contrast, the plasmonic resonance in the reduced state (blue curve) is red-shifted by approximately 85 nm to a centroid wavelength around $\lambda_{c,R}$ = 845 nm. Overall, this matches the spectral shift expected from the refractive index shift of PEDOT [42,44]. Furthermore, a spectral transmission peak around $\lambda$ = 720 nm is found in the reduced state. This peak originates from the overlay of the plasmonic resonance dip at $\lambda_{c,R}$ = 845 nm and the intrinsic PEDOT material transmission dip (absorption peak) around $\lambda$ = 600 nm. This material resonance is also visible in the transmission spectra of only the electropolymerized PEDOT layer in the oxidized and reduced state, which can be found in Figure S3 in the Supplementary Information. The temporal behavior of the centroid wavelength during cyclic voltammetry is depicted in Figure 3b. The upper graph shows the centroid wavelength versus time for a total of four cycles. The centroid wavelengths are determined from the spectral responses of the metasurface from Figure 3a. Centroids of the metasurface in the reduced ($\lambda_{c,R}$) and oxidized state ($\lambda_{c,O}$) are marked in blue and red, respectively. The lower graph in Figure 3b depicts the voltage cycling between -1 V and +0.5 V with a scan rate of 20 mV/s. We find that the centroid wavelength cycles between the oxidized and reduced state and that the PEDOT and thus the hybrid metasurface responds immediately to changes in the voltage. The reduced and oxidized states are separated from each other which allows for stable cycling between two distinct optical states. Furthermore, the switching is reproducible over several cycles with no indication of any degradation.

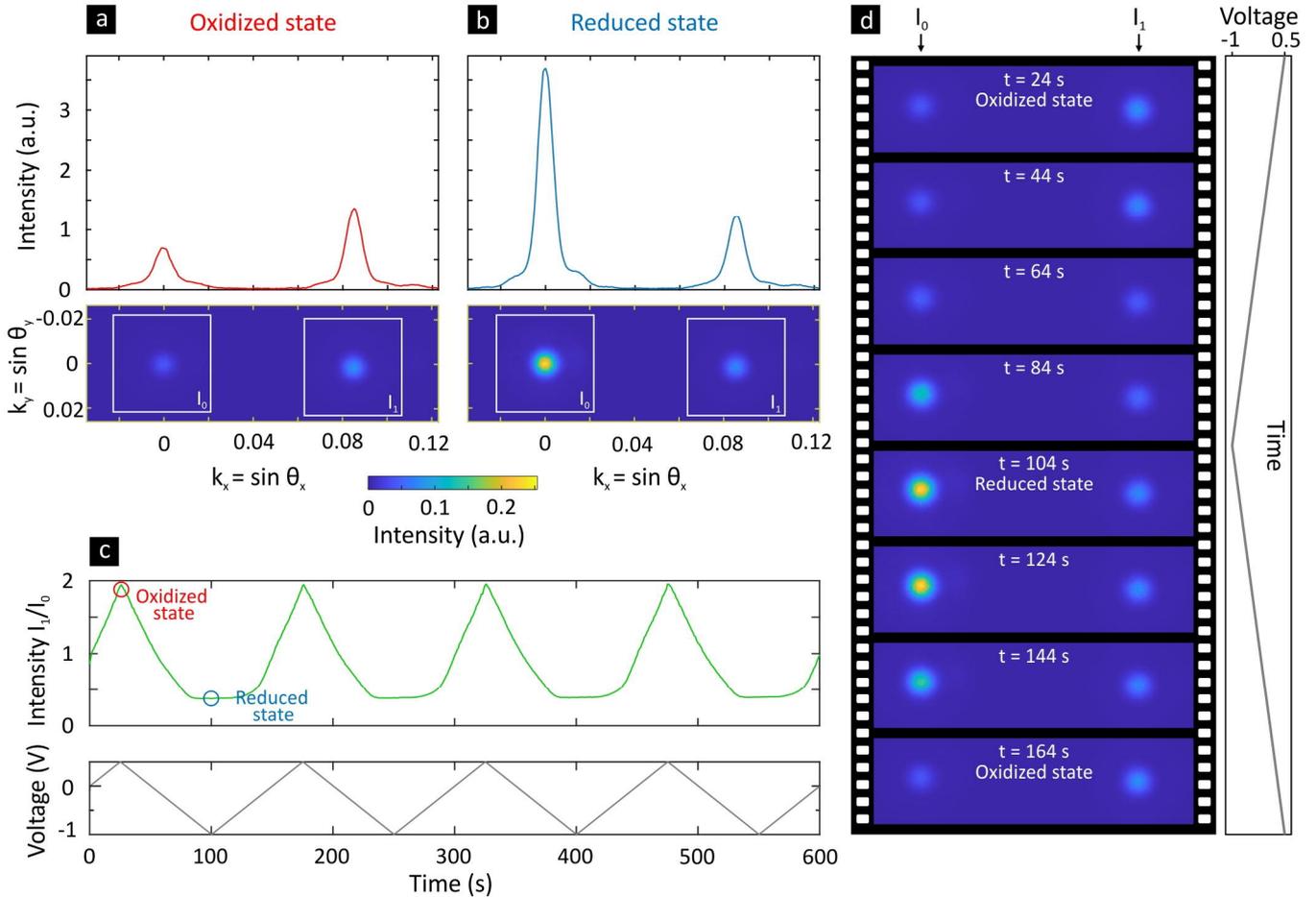

Figure 4. (a) and (b) Intensity on CCD camera in k-space (lower image panels) and integrated intensity profiles (upper panels) for reduced and oxidized state, respectively, displaying $0^{th}$ order beam and $1^{st}$ order diffracted beam. (c) Contrast (intensity ratio $I_1/I_0$ of the first and zeroth order beams) during cyclic voltammetry (tuning of applied voltage). The areas of the respective beam sizes for the intensity calculation are shown with white frames in (a) and (b). (d) Intensity on CCD camera during one cycle. A video showing all frames can be found in the Supplementary Information. The illumination wavelength is $\lambda$ = 750 nm.

As our hybrid metasurface is designed to allow for switchable beam steering in the visible spectral range, we now turn our attention to angle-resolved k-space imaging, as it is illustrated in Figure 4. The beam profiles and intensities of the $0^{th}$ order main and $1^{st}$ order diffracted beam in k-space are illustrated in Figure 4a and b for the oxidized and reduced state, respectively (illumination wavelength $\lambda$ = 750 nm). The lower graphs depict the spatially resolved intensity in k-space, whereas the upper graphs show the corresponding integrated intensities. Overall, we find a gaussian beam profile of both beams. Please note that the intensities are obtained by converting the CCD camera images from sRGB to linear RGB. In the oxidized state in Figure 4a, the diffracted beam $I_1$ has a higher intensity in comparison to the main beam $I_0$. In contrast, this relation inverts for the reduced state in Figure 4b, as here the diffracted beam has a lower intensity than the main beam. We find that the total intensity of both beams increases, which results from the overall higher transmittance of the hybrid metasurface in the reduced state at λ = 750 nm (compare Figure 2a). The diffraction of this designed hybrid metasurface in k-space is $k_x$ = sin $θ_x$ = 0.085, which corresponds to a diffraction angle of $θ_x$ = 4.9°. By changing the superperiod of the metasurface we can increase the diffraction angle to $θ_x$ ≈ 10°. The k-space and temporal response in dependence of the applied voltage of this modified metasurface are shown in Figure S4 in the Supplementary Information.

For potential applications as light modulators, it is important to determine the modulation efficiency of our metasurface. In detail, we determine two quantities necessary to quantify our metasurface relative to existing literature. The first parameter, the basic efficiency

$$\eta^{ox} = \frac{I_1^{ox}}{I_0^{ox}/a}$$

is calculated via the intensity ratio of the diffracted beam $I_1^{ox}$ and the main beam $I_0^{ox}$ in the oxidized PEDOT state. Please note that we introduce and experimentally determine the attenuation coefficient a, which originates from the attenuation of the main beam $I_0$ via the QWP (see Figure 2a). Consequently, we obtain efficiency values of the metasurface in Figure 4 of $\eta^{ox}$ = 3.67%. The efficiency of the modified metasurface in Figure S4 is $\eta^{ox}$ = 0.83%.

A second quantity to compare is the figure of merit [46]

$$FOM = \frac{1}{2}(C^{ox} - C^{red}) = \frac{1}{2}\left(\frac{I_0^{ox}/a - I_1^{ox}}{I_0^{ox}/a + I_1^{ox}} - \frac{I_0^{red}/a - I_1^{red}}{I_0^{red}/a + I_1^{red}}\right),$$

where $I_1^{ox}, I_1^{red}$, and $I_0^{ox}, I_0^{red}$ are the intensities of the diffracted and main beam in the oxidized and reduced state, respectively. $C^{ox}$ and $C^{red}$ denote the contrasts in the oxidized and reduced state, respectively. A complete transfer of the intensity from the main beam to the diffracted beam during switching would result in a $FOM = 1$. For our metasurface in Figure 4, we obtain a value of $FOM = 0.05$, whereas for the modified metasurface in Figure S4 we obtain $FOM = 0.01$. These values are small compared to other active metasurface designs in literature [46], which most likely originates from the comparably small refractive index shift of PEDOT in the visible spectral range.

The intensity ratio $I_1/I_0$ of our initial metasurface during cyclic voltammetry is plotted in Figure 4c. Please note that due to the reduced intensity of the main beam $I_0$, the absolute value of the intensity ratio $I_1/I_0$ is in fact smaller as discussed above for the calculation of the efficiency of the metasurface. A further increase of the intensity ratio and thus diffraction efficiency might be possible by considering other metasurface designs with modified design parameters (metasurface material, antenna geometry, operating wavelength, variation of active material, and many more). Again, we find that the hybrid metasurface quickly reacts to voltage changes. In the oxidized state after t = 24 s, the ratio has a sharp maximum whereas in the reduced state after t = 104 s we find a flat plateau around the minimum. The observed temporal behavior of the intensity ratio clearly shows that we can actively control the intensity of both beams. The intensities of the individual beams during cyclic voltammetry are plotted in Figure S5 in the Supplementary Information. The variations in the beam intensities do not originate from a simple transmittance change of the PEDOT, which would result in a temporal constant intensity ratio $I_1/I_0$, but rather from a change in the diffracted intensity due to its refractive index change and subsequent tuning of the metasurface plasmonic resonance. We find from the individual beam intensities in Figure S5 that it is, in fact, possible to vary the diffracted beam intensity individually while the intensity of the main beam remains constant. Thus, the "gained" intensity in the diffracted beam needs to result in a reduced intensity in higher-order beams and background (mostly diffuse scattering). Consequently, we are able to vary the diffraction-efficiency of the hybrid metasurface via the applied voltage. Selected k-space images of this beam steering process are depicted in Figure 4d. A video showing all frames can be found in the Supplementary Information. The positions of the main ($I_0$) and diffracted beams ($I_1$) are marked with arrows. We observe the switching behavior of the metasurface cycling between a higher intensity in the diffracted beam and a higher intensity in the main beam.

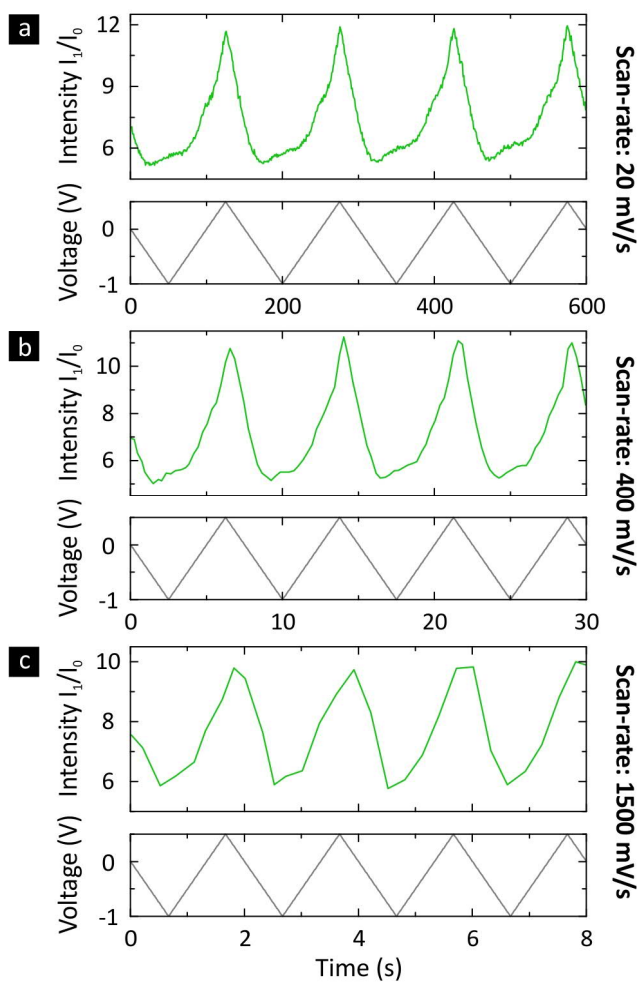

Figure 5. Response of the metasurface for different scan-rates of (a) 20 mV/s, (b) 400 mV/s, and (c) 1500 mV/s. The upper graphs depict the ratio $I_1/I_0$ of the intensity of the diffracted beam $I_1$ and the main beam $I_0$ whereas the lower graphs indicate the applied voltage over time. The response of the switchable metasurface is only limited by the sensitivity and thus frame rate of the used camera. In principle, even faster scan-rates are possible with PEDOT to reach typical display frequencies in the tens of Hertz region.

For practical applications in active and zoomable lens designs or future display technologies, active metasurfaces ideally should have switching speeds at typical display frequencies in the tens of Hz range. Consequently, we investigate as a final step the response and switching time of our hybrid metasurface by varying the scan-rate during cyclic voltammetry. The results are depicted in Figure 5 for scan-rates of (a) 20 mV/s, (b) 400 mV/s, and (c) 1500 mV/s. Please note that we use for these measurements a newly fabricated hybrid metasurface with nominally identical parameters compared to the one used for the characterization in Figure 4. For all measurements in Figure 5, we use linear ramps in constant cyclic phases as shown in the lower graphs of each panel and an exposure time of the CCD camera of 120 ms. For the slowest scan rate of 20 mV/s in Figure 5a, we find that the temporal resolution is high enough to resolve even slightest differences in the intensity ratio $I_1/I_0$ of the diffracted beam $I_1$ and the main beam $I_0$. The peaks in the oxidized state (compare Figure 4c) are sharp and well defined and the performance of the metasurface is reproducible over several cycles with only marginal increases in the overall intensity ratio and no noticeable degradation of the PEDOT layer.

When increasing the scan rate to 400 mV/s we see in Figure 5b that, as expected, we lose temporal resolution in the response curve of the hybrid metasurface which becomes even more prominent for a scan rate of 1500 mV/s. Here, we switch between oxidized and reduced state of PEDOT within 1 s which results in a period of 2 s or a frequency of 0.5 Hz. At this scan rate, the individual data points are far spaced and become clearly visible in the intensity ratio curve. Please note that the time resolution in our experiment is set by the exposure time of the CCD camera. The individual data points in the upper panel of Figure 5c are temporarily spaced by ≈ 200 – 300 ms, which corresponds to the sum of the exposure time of 120 ms and a short time for the software to store each individually captured image. Nevertheless, the hybrid metasurface still oscillates/cycles reproducibly between its oxidized and reduced state. Consequently, the switching speed is limited in our experiment by the sensitivity and thus exposure time and frame rate of our used CCD camera and is not intrinsically limited by the switching speed of the PEDOT layer. In fact, an extension of the switching speeds of our hybrid metasurface to typical display frequencies (tens of Hz range) should be possible by exchanging the cyclic voltammetry with a direct setting (step-function) of applied voltages of +0.5 V and -1 V. Furthermore, several other factors contribute to define the switching speed, which is discussed in the Supplementary Information.

Conclusion

In conclusion, we have demonstrated a novel approach for an optically active as well as externally and electrically switchable hybrid metasurface. We presented a unique combination of a metasurface comprising gold nanoantennas and electropolymerized PEDOT which allowed for an electrochemically activated switchable beam steering. The feasibility of this combination for potential nanophotonic applications was demonstrated via a detailed investigation of the optical and temporal properties of the hybrid metasurface. We used Fourier-space imaging to reveal switching and diffraction angles of up to 10° with excellent conservation of beam profiles in the first diffraction order. A temporal investigation of the intensities of main and diffracted beams showed that we are able to actively control the efficiency of the metasurface and thus the intensity of the diffracted and primary light. We reach switching frequencies around 1 Hz while the extension to display frequencies is only limited by the measurement components and not intrinsically by the optical and electrical properties of our hybrid metasurface. Overall, our approach finds immediate implication in the design, fabrication, and realization of optically active nanophotonic systems that are electrically switchable. Our results will help to develop future optical technologies such as virtual and augmented reality as well as dynamic holography.

Acknowledgements

We acknowledge financial support from the European Research Council (ERC Advanced Grant Complexplas), Bundesministerium für Bildung und Forschung, Deutsche Forschungsgemeinschaft (SPP1839 Tailored Disorder and SPP1391 Ultrafast Nanooptics), and IQST (Integrated Quantum Science and Technology). This project has received funding from the European Research Council (ERC) under the European Union's Horizon 2020 research and innovation programme (grant agreement No 724306).

Disclosures

The authors declare no conflicts of interest.

# Supplementary Information

## for

ELECTRICALLY SWITCHABLE METASURFACE FOR BEAM STEERING USING PEDOT

*Juliane Ratzsch[1], Julian Karst[1], Jinglin Fu[1], Monika Ubl[1], Tobias Pohl[1], Florian Sterl[1], Claudia Malacrida[2], Matthias Wieland[2], Bernhard Reineke[3], Thomas Zentgraf[3], Sabine Ludwigs[2], Mario Hentschel[1], and Harald Giessen[1]*

[1] 4th Physics Institute and Research Center SCoPE, University of Stuttgart, Pfaffenwaldring 57, 70569 Stuttgart, Germany

[2] IPOC-Functional Polymers, Institute of Polymer Chemistry, University of Stuttgart, Pfaffenwaldring 55, 70569 Stuttgart, Germany

[3] Department of Physics, Paderborn University, 33098 Paderborn, Germany

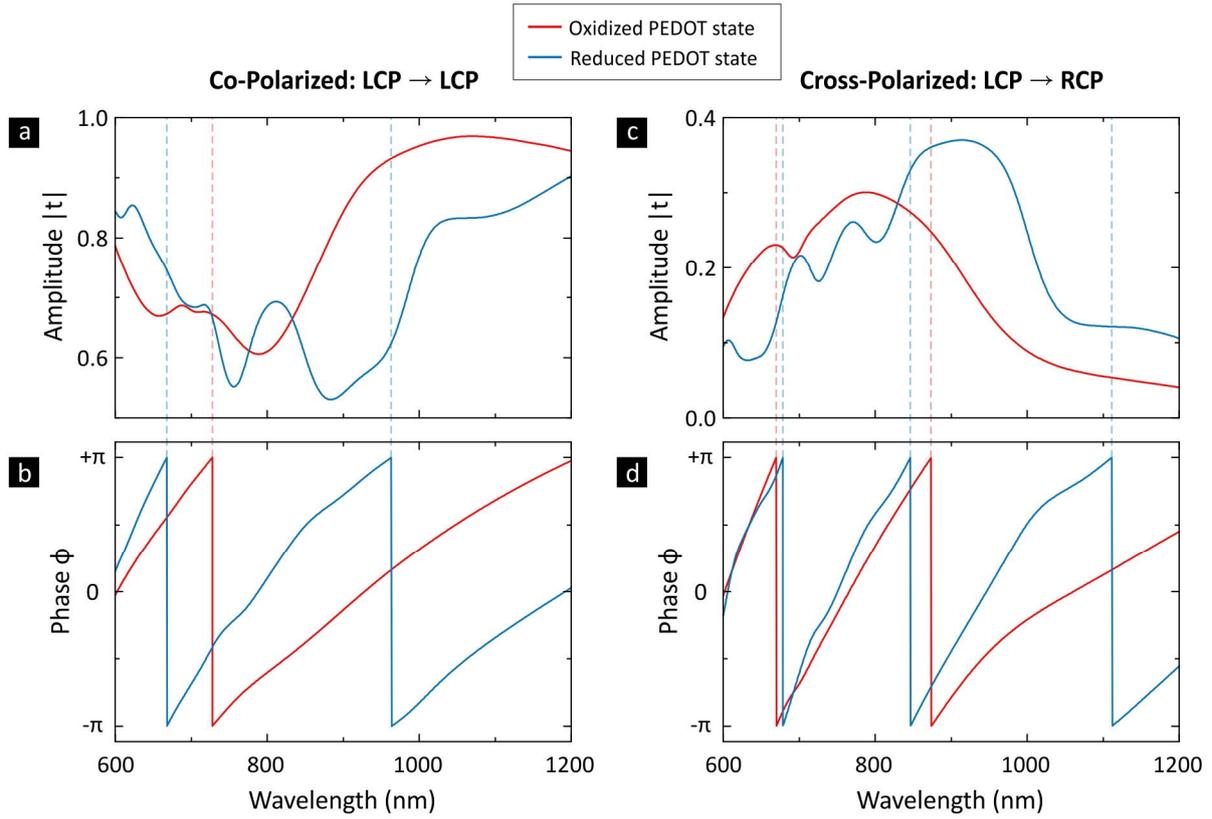

Figure S1. Amplitude $|t|$ and phase $\phi$ of the transmission coefficient $t = |t|e^{i\phi}$ of the simulated metasurface shown in Figure 1 in the main manuscript for PEDOT in the oxidized (red) and reduced state (blue). (a) and (b) depict the amplitude and phase spectra, respectively, for co-polarization (in: LCP, out: LCP). (c) and (d) show the equivalent for cross-polarization (in: LCP, out: RCP). Please note that the other polarization pairs (RCP → RCP, RCP → LCP) are not plotted, as the metasurface is geometrically achiral in 3D. Thus, these combinations are identical to the already shown co-polarized and cross-polarized spectra.

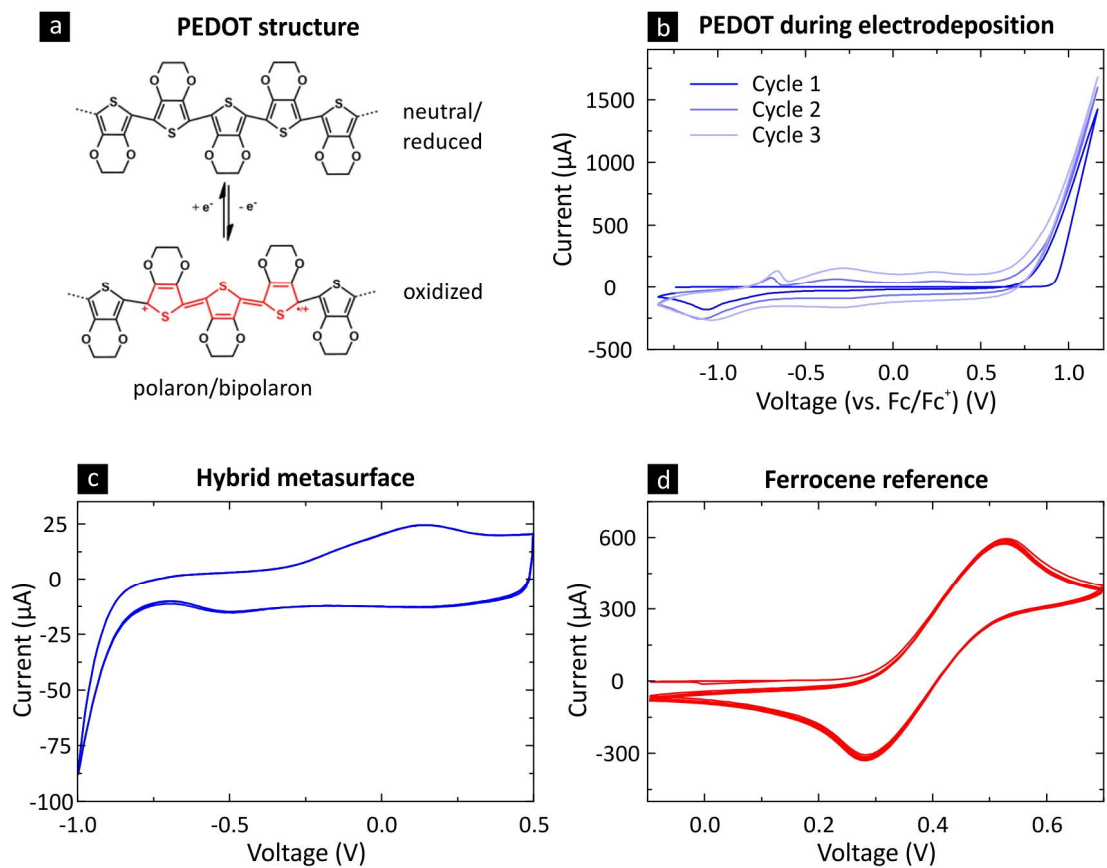

Figure S2. (a) Structures of PEDOT in the neutral/reduced state (top) and in the oxidized state (bottom). (b) CV registered during PEDOT electrodeposition on the hybrid metasurface. The electrodeposition of PEDOT was performed in an air-tight glass cell filled with argon employing a three electrode setup; a Pt plate was used as counter electrode, an AgCl coated Ag wire was used as pseudoreference which was calibrated to the internal standard Fc/Fc$^+$. All samples were deposited by three subsequent voltammetric cycles of a 0.01 mol/L EDOT monomer solution in 0.1 mol/L $CH_3CN$/$Bu_4NPF_6$ within the potential range -1.34 V and 1.16 V vs Fc/Fc$^+$ (-1 V to 1.5 V vs Ag/AgCl; Fc/Fc$^+$ 0.34 V) with a scan-rate of 200 mV/s. The electrodeposition was performed using an Autolab PGSTAT101 and the data were recorded on a computer endowed with the software nova 1.1. (c) Characteristic voltammogram of electrodeposited PEDOT on the hybrid metasurface registered in 0.1 mol/L $CH_3CN$/$Bu_4NPF_6$ with a scan-rate of 20 mV/s. 4 cycles are shown, scan direction from 0 V to 0.5 V to -1 V. Measured vs. Ag/AgCl. (d) CV registered for the internal reference Fc/Fc$^+$ couple.

.

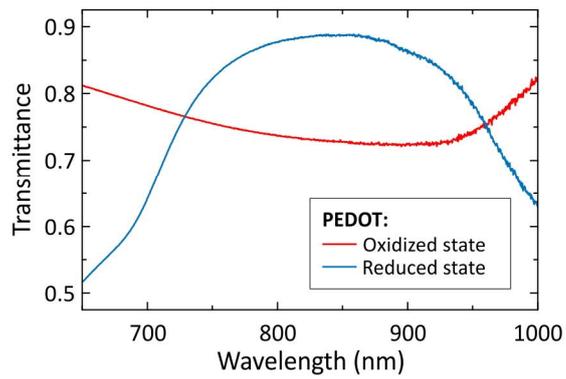

Figure S3. Transmittance spectra of the electropolymerized PEDOT layer in the oxidized (red) and reduced state (blue). We see the onset of the well-known absorption peak (here, transmittance dip) of PEDOT in the reduced state around λ = 600 nm (Wieland M, et al. 2020 *Flex. Print. Electron.* 5 014016).

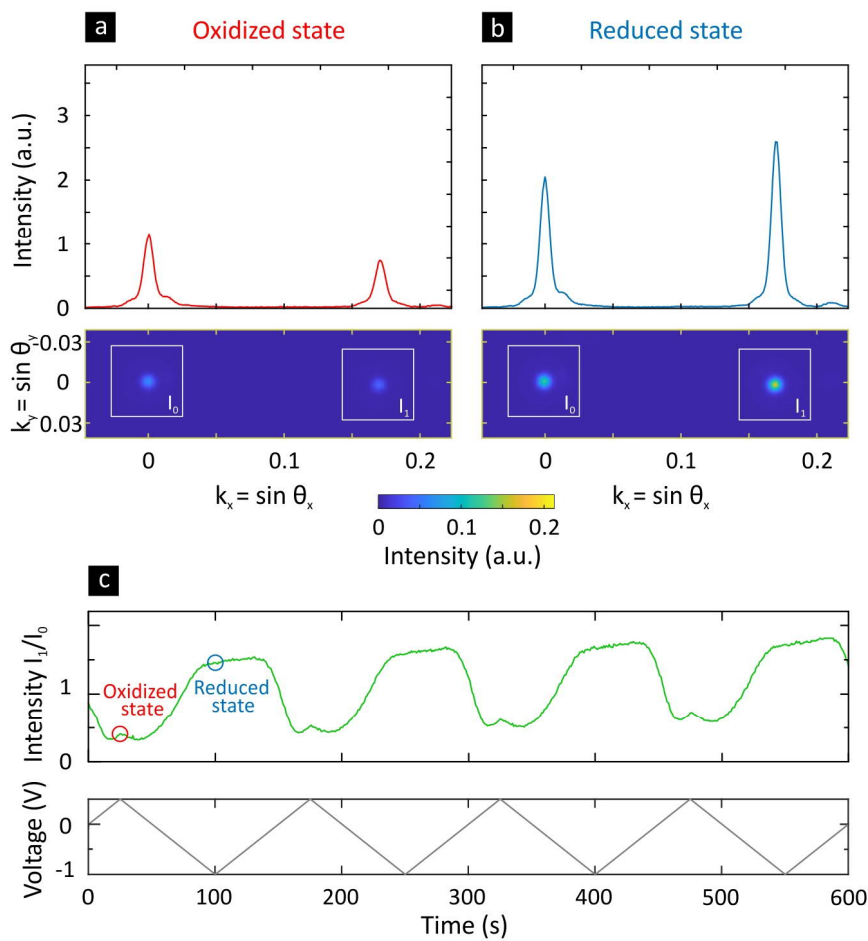

Figure S4. Switching performance of metasurface with changed geometric properties to increase the diffraction angle to ≈ 10°. The antennas have an individual size of 130 nm x 80 nm, a thickness of 50 nm, and a period of 300 nm in x- and y-direction. Adjacent antennas are rotated with an incremental rotation angle of 12°. The resulting superperiod after 180° rotation is 4.5 µm and comprises 15 antennas. (a) and (b) Intensity on CCD camera in k-space (lower image panels) and integrated intensity profiles (upper panels) for reduced and oxidized state, respectively, showing 0$^{th}$ order beam and 1$^{st}$ order diffracted beam. (c) Contrast (intensity ratio $I_1/I_0$) during cyclic voltammetry (tuning of applied voltage). The areas of the respective beam sizes for the intensity calculation are shown with white frames in (a) and (b). A video showing all frames during CV can be found in the Supplementary Information. The illumination wavelength is λ = 750 nm.

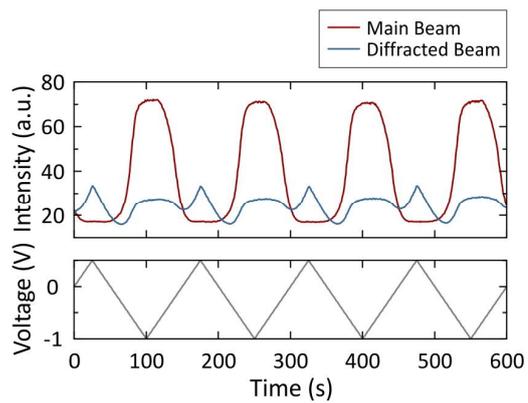

Figure S5. Individual intensities of main (red) and diffracted (blue) beam of the hybrid metasurface shown in Figure 4 in the main manuscript during cyclic voltammetry. We find that the intensity of the diffracted beam can be actively varied via the external voltage (see first onset of blue curve) while the main beam remains constant.

## 1.1. Increase of Switching Speeds

Switching time is the time necessary for the coloring/bleaching process of an electrochromic material. Several factors contribute to define the switching speed, these include for example the magnitude of the applied potential, the ionic conductivity of the electrolyte, ion diffusion in thin films which is itself related to film thickness and morphology. Indeed, one straightforward way to increase switching-speeds is the employment of very thin films characterized by absence of diffusion limitations as the following processes are determining the charging/discharging process of a conducting polymer film: electron exchange reaction (hopping), percolation, and counterion diffusion.[1]–[3] In general, a variation in the doping level of the polymer film is directly associated with a variation in its absorption properties; general relationships between electrochemical dimensions have been determined for different electrochemical techniques.[1,4]

A common way of measuring switching time is by square-wave potential steps applied to the film (or device) with concurrent analysis of transmittance variation at a single wavelength with determination of the time required to obtain a certain percentage of color switch. Similar analysis could be performed by determining current variations also by a desired percentage.[5]